\documentclass[a4paper,10pt,twocolumn]{article}

\usepackage[english]{babel}   

\usepackage{amsmath}
\usepackage{graphicx}
\usepackage{epstopdf}
\usepackage{color}
\usepackage[usenames,dvipsnames]{xcolor}
\usepackage{listings}
\usepackage[utf8]{inputenc}
\usepackage{url}
\usepackage{hyperref}
\usepackage{xspace}
\usepackage{subfigure}
\usepackage{listings}
\usepackage{float}
\def\draft#1{{#1}}  
\def\draft#1{}   

\newcommand{\todo}[1]{}
\newcommand{\code}[1]{\texttt{#1}}

\newcommand{\proj}{Chemora\xspace}
\newcommand{\term}[1]{{\sl#1}}
\newcommand{\param}[1]{{\tt#1}}
\long\def\bitbucket#1{}

\begin{document}

\title{From Physics Model to Results: An Optimizing Framework for Cross-Architecture Code Generation
\author{Marek Blazewicz$^{1,2,*}$, Ian Hinder$^{3}$, David M. Koppelman$^{4,5}$, 
  Steven R. Brandt$^{4,6}$,\\ 
  Milosz Ciznicki$^{1}$, Michal Kierzynka$^{1,2}$, Frank Löffler$^{4}$, 
  Erik Schnetter$^{7,8,4}$, Jian Tao$^{4}$ 
  \let\thefootnote\relax\footnote{$^{1}$ Applications Department,
    Pozna\'{n} Supercomputing \& Networking Center, Pozna\'{n}, Poland}
  \let\thefootnote\relax\footnote{$^{2}$ Pozna\'{n} University of Technology, 
    Pozna\'{n}, Poland}
  \let\thefootnote\relax\footnote{$^{3}$ Max-Planck-Institut für
    Gravitationsphysik, Albert-Einstein-Institut, Postdam, Germany}
  \let\thefootnote\relax\footnote{$^{4}$ Center for Computation \&
    Technology, Louisiana State University, Baton Rouge, LA, USA}
  \let\thefootnote\relax\footnote{$^{5}$ Division of Electrical \&
    Computer Engineering, Louisiana State University, Baton Rouge, LA, USA}
  \let\thefootnote\relax\footnote{$^{6}$ Division of Computer
    Science, Louisiana State University, Baton Rouge, LA, USA}
  \let\thefootnote\relax\footnote{$^{7}$ Perimeter Institute for
    Theoretical Physics, Waterloo, ON, Canada}
  \let\thefootnote\relax\footnote{$^{8}$ Department of Physics,
    University of Guelph, Guelph, ON, Canada}
  \let\thefootnote\relax\footnote{$^{*}$ Email: marqs@man.poznan.pl}
}
}

\date{29th April, 2013}
\maketitle
\draft{\tableofcontents}

\begin{abstract}
Starting from a high-level problem description in terms of
partial differential equations using abstract tensor notation, the \emph{\proj}
framework discretizes, optimizes, and generates complete high performance codes
for a wide range of compute architectures. \proj extends the capabilities of Cactus, facilitating the usage of large-scale CPU/GPU systems in 
an efficient manner for complex applications, without low-level code tuning.
\proj achieves parallelism through
MPI and multi-threading, combining OpenMP and CUDA\@.
Optimizations include high-level code
transformations, efficient loop traversal strategies, dynamically
selected data and instruction cache usage strategies, and JIT
compilation of GPU code tailored to the problem characteristics.
The discretization is based on higher-order finite
differences on multi-block domains.
\proj's capabilities are demonstrated by simulations of black
hole collisions. This problem provides an acid test of the framework,
as the Einstein equations contain hundreds of variables and thousands of
terms.
\end{abstract}

\section{Introduction}
\label{sec:intro}

High performance codes are becoming increasingly difficult to program,
despite a proliferation of successful (but incremental) efforts to
increase programmability and productivity for high performance
computing (HPC) systems. The reasons for this range over several
layers, beginning with the need for large, international
collaborations to combine expertise from many different fields of
science,
to the need to address a wide variety of systems and hardware
architectures to ensure efficiency and performance.

As heterogeneous and hybrid systems are becoming common in HPC
systems, additional levels of parallelism need to be addressed, and
the bar for attaining efficiency is being raised. Three out of ten,
and 62 of the top 500 of the fastest computers in the world use
accelerators of some kind to achieve their performance~\cite{top500}.
More large heterogeneous systems are scheduled to be set up, especially
including new Intel Xeon Phi and Nvidia K20x co-processors.
\todo{ES: Should we update this statement? Stampede at TACC (with Xeon
  Phis) is open to the public, and Blue Waters (with Nvidia GK110
  coprocessors) is open for testing.}

In this paper we present \emph{\proj}, using an integrated approach
addressing programmability and performance at all levels, from
enabling large-scale collaborations, to separating physics, numerical
analysis, and computer science portions, to
disentangling kernel implementations from performance optimization
annotations. \proj is based on the \emph{Cactus}
framework~\cite{Goodale02a, cactusweb}, a well-known tool used in
several scientific communities for developing HPC applications. Cactus
is a component-based framework providing key abstractions to 
significantly simplify parallel programming for a large class of problems, in
particular solving systems of partial differential equations (PDEs) on
block-structured grids -- i.e.\ adaptive mesh refinement (AMR) and
multi-block systems (see section \ref{sec:cactus} below).

\proj enables existing Cactus-based applications to continue scaling
their scientific codes and make efficient use of new hybrid systems,
without requiring costly re-writes of application kernels or adopting
new programming paradigms. At the same time, it also provides a
high-level path for newly developed applications to efficiently employ
cutting-edge hardware architectures, without having to target a
specific architecture.

We wish to emphasize that the present work is merely the next step in
the currently fifteen year-long history of the Cactus framework. While
finding ways to exploit the power of accelerators is perhaps the
largest current challenge to increased code performance, it is really
only the latest advance in an ever-changing evolution of computer
architectures. Suport for new architectures is typically added to the
lower-level components of frameworks (such as Cactus) by the framework
developers, allowing the application scientist to take advantage of
them without having to significantly rewrite code.

To create the \proj framework, we have built on top of a number of
existing modules that have not been written specifically for this
project, as well as creating new modules and abstractions. The main
research and development effort has been the integration of these
modules, especially as regards accelerator interfaces, their
adaptation for production codes as well as automatic optimizations to
handle complicated Numerical Relativity codes. The result is that this
framework allows the use of accelerator hardware in a transparent and
efficient manner, fully integrated with the existing Cactus framework,
where this was not possible before.  The full contribution to the
described research work has been described in the section
\ref{sec:contribution}. The framework, along with introductory
documentation, will be made publicly available \cite{chemoracode}.

\subsection{Scientific Motivation}
\label{sec:science}

Partial differential equations are ubiquitous throughout the fields of
science and engineering, and their numerical solution is a challenge
at the forefront of modern computational science.  In particular, our
application is that of \emph{relativistic astrophysics}. Some of the
most extreme physics in the universe is characterised by small regions
of space containing a large amount of mass, and Newton's theory of
gravity is no longer sufficient; Einstein's theory of General
Relativity (GR) is required.  For example, black holes, neutron stars,
and supernovae are fundamentally relativistic objects, and
understanding these objects is essential to our understanding of the
modern universe. Their
accurate description is only possible using GR\@. The solution of
Einstein's equations of GR using computational techniques is known as
\emph{numerical relativity} (NR\@). See \cite{Pfeiffer:2012pc} for a recent
review, and see \cite{Loffler:2011ay} for a detailed description of an open-source
framework for performing NR simulations.

One of the most challenging applications of NR is the inspiral and
merger of a pair of orbiting black holes.  GR predicts the existence
of gravitational waves: ripples in spacetime that propagate away from
heavy, fast-moving objects.  Although there is indirect evidence,
these waves have not yet been directly detected due to their low
signal strength. The strongest expected sources of gravitational waves are
binary black hole and neutron star mergers, and supernova explosions--
precisely those objects for which GR is required for accurate
modeling.  Several gravitational wave detectors \cite{Fritschel:2003qw} are
presently under construction and they are expected to see a signal within
the next few years.  The detection of gravitational waves will lead
to an entirely new view of the universe, complementary to existing
electromagnetic and particle observations.  The existence and
properties of expected gravitational wave sources will dramatically
extend our knowledge of astronomy and astrophysics.

NR models the orbits of the black holes, the waveforms they produce,
and their interaction with these waves
using the Einstein equations. Typically, these equations are split
into a 3+1 form, breaking the four dimensional character of the
equations and enabling the problem to be expressed as a time evolution
of gravitational fields in three spatial dimensions.
The Einstein equations in the BSSN formulation~\cite{Nakamura:1987zz,
  Shibata:1995we, Baumgarte:1998te}
are a set of coupled nonlinear
partial differential equations with 25 variables~\cite{Alcubierre99d, Alcubierre02a},
usually written for compactness in abstract index form.
When fully expanded, they contain thousands of terms, and the right
hand side requires about 7900
floating point operations per grid point to evaluate once, if using
eigth order finite differences.

The simulations are characterised by the black hole mass, $M$,
a length, $G M/c^2$, and a time, $G M/c^3$. Usually one 
uses units in which $G = c = 1$, allowing both time and distance to be
measured by $M$. Typical simulations of the type listed above
have gravitational waves of
size $\sim 10 M$, and the domain to be simulated is $\sim
100$--$1000 M$ in radius.  For this reason, Adaptive Mesh Refinement
(AMR) or multi-block methods are required to perform long-term BBH
simulations.

Over 30 years of research in NR culminated in a major breakthrough in
2005~\cite{pretorius2005evolution,Baker:2005vv,Campanelli:2005dd},
when the first successful long-term stable binary black hole
evolutions were performed.  Since then, the NR community has refined
and optimized their codes and techniques, and now routinely runs
binary black hole simulations, each employing hundreds or thousands of
CPU cores simultaneously of
the world's fastest supercomputers.
Performance of the codes is a critical issue, as the
scientific need for long waveforms with high accuracy is compelling.
One of the motivations of the \proj project was taking the NR
codes into the era of computing with the use of accelerators (in particular
GPUs) and improving their performance by an order of magnitude, thus enabling
new science.

\subsection{Related Work}

To achieve sustained performance on hybrid supercomputers and reduce
programming cost, various programming frameworks and tools have been developed, e.g.,
Merge~\cite{Linderman:2008:MPM:1353536.1346318} (a library based framework
for heterogeneous multi-core systems), Zippy~\cite{CGF:CGF1131} (a framework
for parallel execution of codes on multiple GPUs),
BSGP~\cite{Hou:2008:BBG:1360612.1360618} (a new programming language for
general purpose computation on the GPU), and
CUDA-lite~\cite{springerlink:10.1007/978-3-540-89740-8_1} (an enhancement to
CUDA that transforms code based on annotations).
Efforts are also underway to improve compiler tools
for automatic parallelization and optimization of affine loop
nests for GPUs~\cite{Baskaran:2008:CFO:1375527.1375562} and for automatic
translation of OpenMP parallelized codes to
CUDA~\cite{Lee:2009:OGC:1594835.1504194}.
Finally, OpenACC is slated to provide OpenMP-like annotations for C and
Fortran code.

Stencil computations form the kernel of many scientific applications that 
use structured grids to solve partial differential equations.
This numerical problem can be characterised as the {\em structured grids} "Berkeley
Dwarf" \cite{berkeleydwarfs2006}, one of a set of algorithmic patterns identified as important
for current and near-future computation.
In particular, stencil computations parallelized using hybrid architectures
(especially multi-GPU) are
of particular interest to many researchers who want to leverage the emerging hybrid
systems to speed up scientific discoveries.
Micik~\cite{Micik2009} proposed an optimal 3D finite difference
discretization of the wave equation in a CUDA environment, and
also proposed a way to minimize the latency of inter-node communication
by overlapping slow PCI-Express (interconnecting the GPU with the
host) data exchange with computations. This may be achieved by
dividing the computational domain along the slowest varying dimension.
Thibault \cite{Thibault2009} followed the idea of a domain division pattern and implemented
a 3D CFD model based on finite-difference discretization of the Navier-Stokes equations parallelized
on a single computational node with 4 GPUs.

Jacobsen \cite{Jacobsen2010} extended this model by adding inter-node communication via
MPI\@. They followed the approach described in Micik~\cite{Micik2009} and overlapped the communication with
computations as well as GPU-host with host-host data exchange. However, they did not take
advantage of the full-duplex nature of the PCI-Express bus, which would have decreased the
time spent for communication. Their computational model also divides the domain along the slowest
varying dimension only, and this approach is not suitable for all numerical problems. For example, for large computational
domains, the size of the ghost zone becomes noticeable in comparison to the computed part
of the domain, and the communication cost becomes larger than the computational cost, which can
be observed in the non-linear scaling of their model.

Notable work on an example stencil application was selected as a finalist of the Gordon Bell Prize in 
SC 2011 as the first peta-scale result \cite{Shimokawabe2011}. Shimokawabe et al.\ demonstrated very high 
performance of 1.017 PFlop/s in single precision using 4,000 GPUs along with 16,000 CPU cores on TSUBAME 2.0. 
Nevertheless, a set of new and more advanced optimization techniques introduced in the \proj framework as 
well as its capabilities to generate highly efficient multi-GPU stencil computing codes from a high-level 
problem description make this framework even more attractive for users of large-scale hybrid systems.

Physis \cite{Physis} addresses the problem of dividing the domain in
all dimensions, and is these days
seen as one of the most efficient frameworks for stencil
computations over regular multidimensional Cartesian
grids in distributed memory environments.
The framework in its current state, however, does not divide
the domain automatically; this has to be done manually
at launch time.
Nevertheless, Physis achieves very good scaling by taking
advantage of memory transfers overlapped with computations.
Stencil computations are defined in the form of C-based functions (or \emph{kernels})
with the addition of a few special macros that allow accessing values at grid points.
The framework also uses CUDA streams that allow for parallel execution
of multiple kernels at the same time; e.g.\ regular and boundary kernels
may be executed in parallel.
Data dependencies between stencil points are resolved statically,
hence must be known beforehand, at compile time.
The authors put a special emphasis on ease of use, and
indeed the time needed to write an application in Physis is relatively short.
This framework was evaluated using three benchmark programs running on 
the TSUBAME~2.0 supercomputer, and proved
to generate scalable code for up to 256 GPUs.
Below, we compare \proj with its dynamic compilation and 
auto-tuning methods to Physis, and show that \proj outperforms Physis
in the area of automatically generated
code for GPU clusters.

\subsection{Contributions}
\label{sec:contribution}
This paper makes the following contributions:

\begin{itemize}
\setlength{\itemsep}{-2pt}
\item An overview of the \proj framework for generating hybrid
  CPU\slash GPU cluster code from PDE descriptions is presented and
  its performance is characterized.

\item A language for expressing differential equation models of
  physical systems suitable for generating hybrid cluster simulation
  code (based on the existing \term{Kranc} code-generation package), was developed.

\item Model-based GPU tile\slash thread configuration optimization
  techniques were developed, enabling the exploration of a large
  search space and the use of dynamic compilation (performed once on
  the chosen configuration).

\item Automatic hybrid execution GPU\slash CPU data staging techniques
  were developed (the \term{accelerator} module).

\item GPU tuning techniques were developed for large kernel codes,
  such as register-pressure sensitive configuration.

\item The first demonstration binary black hole simulations using GPUs in full GR
  were presented.  Since Chemora has not yet been applied to the
  Carpet AMR driver, these are not suitable for production physics,
  but prove that existing codes used in numerical relativity can be
  adapted to Chemora.
\end{itemize}

\section{\proj Framework}

\begin{figure}
\centering
\includegraphics[width=0.5\textwidth]{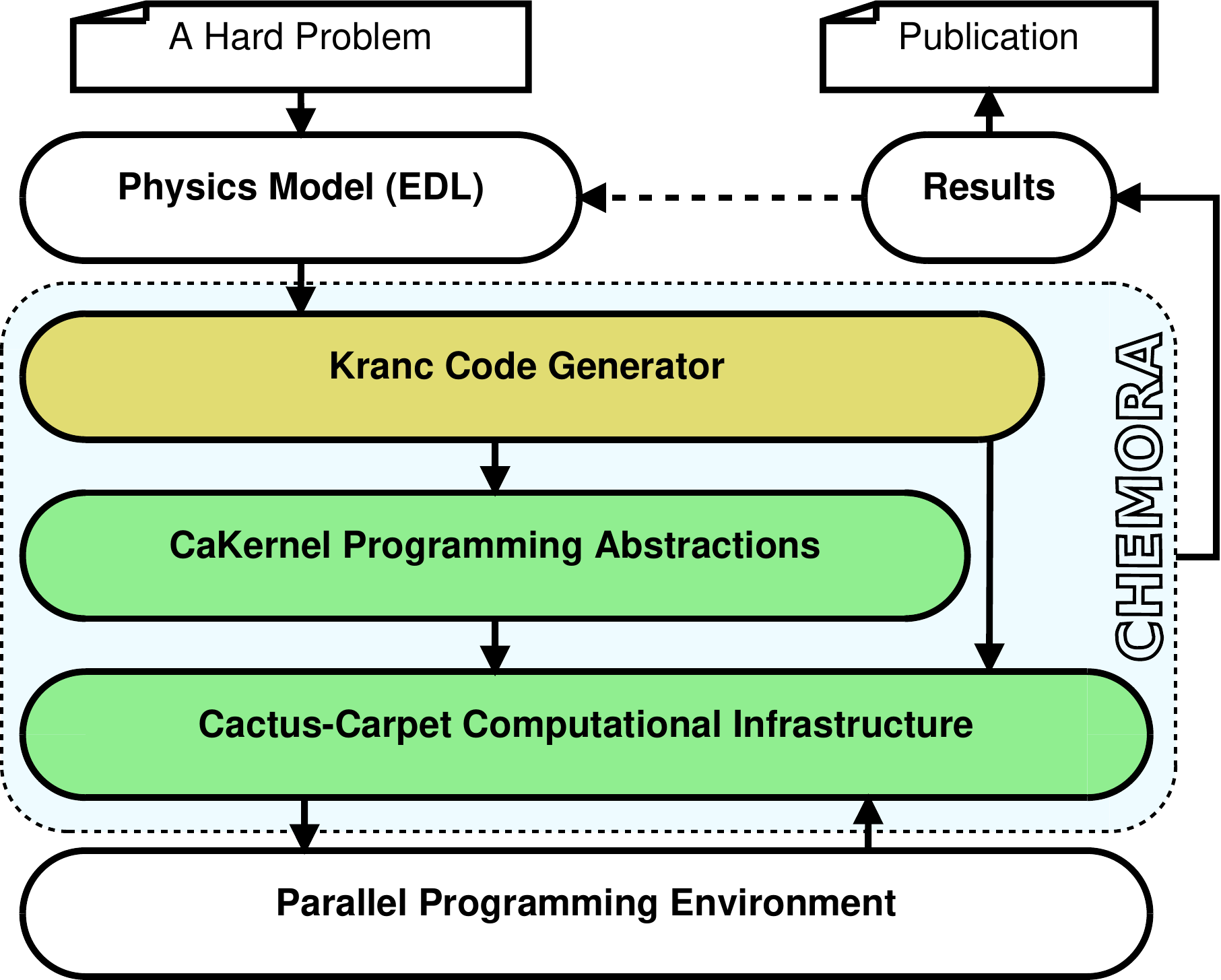}
\caption{An architectural view of \proj. \proj consists of three major
components: The Cactus-Carpet computational infrastructure, CaKernel
programming abstractions, and the Kranc code generator. \proj takes a physics
model described in a high level Equation Description Language and 
produces highly optimized code suitable for parallel execution on 
heterogeneous systems.}
\label{fig:chemora_arch}
\end{figure}

\proj takes a physics model described in a high level 
\emph{Equation Description Language} (EDL) and generates highly optimized code suitable 
for parallel execution on heterogeneous systems.
There are three major components in \proj:
the Cactus-Carpet computational infrastructure, 
CaKernel programming abstractions, and the Kranc
code generator. \proj is portable to many
operating systems, and adopts widely-used parallel programming 
standards (MPI, OpenMP and OpenCL) and models (vectorization and CUDA\@).
An architectural view of the \proj framework is shown in
Figure~\ref{fig:chemora_arch}. We describe the individual components below.

\subsection{Cactus-Carpet Computational Infrastructure}
\label{sec:cactus}

The Cactus computational framework is the foundation of \proj.
Cactus~\cite{Goodale02a, cactusweb} is an open-source, modular,
highly-portable programming environment for collaborative research
using high-performance computing. Cactus is distributed with a generic
computational toolkit providing parallelization, domain decomposition,
coordinates, boundary conditions, interpolators, reduction operators,
and efficient I/O in different data formats. More than 30
groups worldwide are using Cactus for their research work in cosmology,
astrophysics, computational fluid dynamics, coastal modeling, quantum
gravity, etc. The Cactus framework is a vital part of the Einstein
Toolkit~\cite{Loffler:2011ay, EinsteinToolkit:web},
an NSF-funded collaboration enabling a
large part of the world-wide research in numerical relativity by
providing necessary core computational tools as well as a common
platform for exchanging physics modules. Cactus is part of the
software development effort for Blue Waters, and in particular the
Cactus team is working with NCSA to produce development interfaces and
paradigms for large scale simulation development.

One of the features of Cactus relevant in this context is that it
externalizes parallelism and memory management into a module (called
a \emph{driver}) instead of providing it itself,
allowing application modules (called \emph{thorns}) to function mostly
independently of the system architecture. Here we employ the
\emph{Carpet} driver~\cite{Schnetter-etal-03b, Schnetter06a,
  carpetweb} for MPI-based parallelism via spatial domain
decomposition. Carpet provides adaptive mesh refinement (AMR) and
multi-block capabilities\footnote{We do not use these capabilities in
  the examples below.}, and has been shown to scale to more than
16,000 cores on current NERSC and XSEDE systems.

In the typical Cactus programming style for application modules, these
modules consist either of \emph{global} routines (e.g.\ reduction or
interpolation routines), or \emph{local} routines (e.g.\ finite
differencing kernels). Local routines are provided in the form of
kernels that are mapped by the driver onto the available resources.
At run time, a schedule is constructed, where Cactus orchestrates the
execution of routines as well as the necessary data movement
(e.g.\ between different MPI processes).
This execution model is both easy to understand for application
scientists, and can lead to highly efficient simulations on large
systems. Below, we refine this model to include accelerators
(e.g.\ GPUs) with separate execution cores and memory systems.

\subsection{CaKernel Programming Abstractions}
\label{sec:cakernel}
The \proj programming framework uses the CaKernel
\cite{parco11, ppopp11, sciprog11}, a set of high level programming
abstractions, and the corresponding implementations.
Based on the Cactus-Carpet computational infrastructure,
CaKernel provides two major sets of programming abstractions:
(1) \emph{Grid Abstractions} that represent the dynamically 
distributed adaptive grid
hierarchy and help to separate the application development from the
distributed computational domain;
(2) \emph{Kernel Abstractions} that enable automatic generation of numerical
kernels from a set of highly optimized templates and help to separate the
development, scheduling, and execution of numerical kernels.

\subsubsection{Grid Abstractions}
The Cactus flesh and the Cactus computational toolkit contain a collection
of data structures and functions that 
can be categorized into the following three grid abstractions, which commonly appear
in high level programming frameworks for parallel block-structured
applications~\cite{parabrow96}:
\begin{itemize}
\setlength{\itemsep}{-2pt}
\item The \emph{Grid Hierarchy (GH)} represents the distributed adaptive GH\@.
The abstraction enables application developers to create, operate and destroy
hierarchical grid structures. The regridding and partitioning operations on a grid
structure are done automatically whenever necessary. In Cactus, grid operations
are handled by a driver thorn which is a special module in Cactus.
\item A \emph{Grid Function (GF)} represents a distributed data structure
containing one of the variables in an application. Storage, synchronization, arithmetic,
and reduction operations are implemented for the GF by standard thorns. The
application developers are responsible for providing routines for
initialization, boundary updates, etc.
\item The \emph{Grid Geometry (GG)} represents the coordinates, bounding boxes,
and bounding box lists of the computational domain. Operations on the GG, such
as union, intersection, refine, and coarsen are usually implemented in a driver
thorn as well.
\end{itemize}

\subsubsection{Kernel Abstractions}
The kernel abstractions enable automatic code generation with a set of highly optimized
templates to simplify code construction. The definition of a kernel requires
the following three components:
\begin{itemize}
\setlength{\itemsep}{-2pt}
\item A \emph{CaKernel Descriptor} describes one or more numerical
  kernels,
  dependencies, such as grid functions and parameters
  required by the kernel, and grid point relations with its neighbors.
  the information provided in the descriptor is then used to generate
  a kernel frame (macros) that performs automatic data fetching,
  caching and synchronization with the host.
\item A \emph{Numerical Kernel} uses kernel-specific auto-generated
  macros. The function may be generated via other packages (such as
  Kranc), and operates point-wise.
\item The \emph{CaKernel Scheduler} schedules CaKernel launchers and
  other CaKernel functions in exactly the same way as other Cactus
  functions. Data dependencies are evaluated and an optimal strategy
  for transferring data and performing computation is selected
  automatically.
\end{itemize}
These kernel abstractions not only enable a simple way to write and execute
numerical kernels in a heterogeneous environment, but also enable lower-level
optimizations without modifying the kernel code itself.

\subsubsection{Hardware Abstraction}
CaKernel provides an abstraction of the hardware architecture, and
Chemora code is generated on top of this abstraction. The high level
problem specification in the Chemora framework may thus remain
independent of the architecture. The support for new architectures is
the responsibility of the Chemora developers, and thus it is
transparent to the end-user, who should not need to significantly
modify their code once the underlying CaKernel implementation has been
modified.

\subsection{Describing a Physics Model}
Programming languages such as C or Fortran offer a very low level of
abstraction compared to the usual mathematical notation. Instead of
requiring physicists to write equations describing PDEs at this level,
we introduce EDL, a
domain-specific language for specifying systems of PDEs as well as
related information (initial and boundary conditions, constraints,
analysis quantities, etc.) EDL allows equations to be specified independent
of their discretization, allows abstract index notation to be used as a
compact way to write vectors and tensors, and does not limit
the options for memory layout or looping order. For \proj, we designed EDL
from scratch instead of piggybacking it onto an existing language
such as Mathematica, Haskell, or C++ so that we could choose a syntax
that is easily understood by domain scientists, i.e.\ physicists and
engineers.

EDL has a very simple syntax, similar to C, but extended with a
LaTeX-like syntax for abstract index notation for vectors and tensors.
Sample \ref{fig:edl} shows as an example the main part of specifying the
scalar wave equation in a fully first order form (assuming, for
simplicity, the propagation speed is $1$.) In addition to specifying
the equations themselves, EDL supports constants, parameters,
coordinates, auxiliary fields, and conditional expressions.

\begin{lstlisting}[escapechar=!, caption={Example showing (part of) 
the scalar wave equation written in \emph{EDL}, a language designed to describe PDEs. A LaTeX-like
  syntax allows a compact notation for vectors and tensors. Additional
  annotations (not shown here) are needed to complete the
  description.}, label=fig:edl]

!\color{ForestGreen}{begin calculation}! !\color{blue}{Init}!
  u   = !\color{cyan}{0}!
  rho = A exp(!\color{cyan}{-1/2}! (r/W)**!\color{cyan}{2}!)
  v_i = !\color{cyan}{0}!
!\color{ForestGreen}{end calculation}!

!\color{ForestGreen}{begin calculation}! !\color{blue}{RHS}!
  D_t u   = rho
  D_t rho = delta^ij D_i v_j
  D_t v_i = D_i rho
!\color{ForestGreen}{end calculation}!

!\color{ForestGreen}{begin calculation}! !\color{blue}{Energy}!
  eps = !\color{cyan}{1/2}! (rho**!\color{cyan}{2}! + delta^ij v_i v_j)
!\color{ForestGreen}{end calculation}!
...
\end{lstlisting}

In addition to describing the system of equations, EDL makes it possible
to specify a particular discretization by specifying sets of finite
differencing stencils.
These
stencil definitions remain independent of the equations themselves.

The \emph{Kranc} code-generation package (see section
\ref{sec:kranc}), written in Mathematica and
described below, has been enhanced in \proj to accept EDL as its input
language.  Via a J/Link interface to the Piraha PEG \cite{brandt2010piraha} Java
parsing library, the EDL is parsed into Mathematica expressions
equivalent to those traditionally used as input to Kranc.  The formal
grammar which defines the syntax of the language is available as part
of the Kranc distribution, should other tools need to parse EDL files.

In spite of its apparent simplicity, the high-level description in EDL
captures everything that is needed to create a complete Cactus module.
Metadata such as variable declarations, schedule items, and parameter
definitions are extracted from EDL, and implementation choices such as
memory layout and loop traversal order are made automatically or even
dynamically at run time (see below).

Kranc is written in Mathematica, and prior to \proj was used by
writing a script in the Mathematica language to set up data structures
containing equations and then call Kranc Mathematica functions to
generate the Cactus module.  This allowed great flexibility, but at
the same time required users to know the Mathematica language, which
in several ways is idiosyncratic and is unfamiliar to many users.
Additionally, the use of an imperative language meant that Kranc was
unable to reason about the input script in any useful manner (for
example for the purpose of reporting line numbers where errors were
found).  A new, simple, declarative domain-specific language was
therefore created which allowed a concise expression of exactly the
information needed by Kranc.  Existing languages familiar to the
majority of scientists (C, Fortran, Perl, Python) introduce a wide
variety of features and semantics unnecessary for our application, and
none of these are suitable for expressing equations in a convenient
manner.  The block structure of EDL was inspired by Fortran, the
expression syntax by C, and the index notation for tensors by LaTeX.
We feel that the language is simple enough that it can be learned very
quickly by reference to examples alone, and that there is not a steep
learning curve.

By providing a high-level abstraction for an application scientist,
the use of EDL substantially reduce the time-to-solution, which includes:
learning the software syntax, development time from a given system of
equations to machine code, its parallelization on a heterogeneous
architecture, and finally its deployment on production clusters.  It
also eliminates many potential sources of errors introduced by low
level language properties, and thus reduces testing time. For further
information about the total time-to-solution, see
\cite{hochstein2005parallel}.  

\subsection{Automated Code Generation with Kranc}
\label{sec:kranc}

Translating equations from a high-level mathematical notation into C
or Fortran and discretizing them manually is a tedious, error-prone
task. While it is straightforward to do for simple algorithms, this
becomes prohibitively expensive for complex systems.
We identify two levels
of abstraction.  The first is between the continuum equations and the
approximate numerical algorithm (discretization), and the second is
between the numerical algorithm and the computational implementation.

We employ \emph{Kranc}~\cite{Husa:2004ip, Lechner:2004cs, Kranc:web}
as a code-generation package which implements these abstractions. The
user of Kranc provides a \emph{Kranc script} containing a section
describing the partial differential equations to solve, and a section
describing the numerical algorithm to use. Kranc translates this
high-level description into a complete Cactus module, including C++ code
implementing the equations using the specified numerical method, as
well as code and metadata for integrating this into the Cactus
framework.

By separating mathematical, numerical, and computational aspects,
Kranc allows users to focus on each of these aspects separately
according to their specialization. Although users can write Kranc
scripts directly in Mathematica, making use of the EDL 
shields them from
the (sometimes arcane) Mathematica syntax (because they are required to follow a
strict pattern for specifying PDEs) and provides them with much more
informative (high-level) error messages.  Either the traditional Mathematica language, or the new EDL language, can be used
with Chemora for GPU code generation. 

Kranc is able to:
\begin{itemize}
\setlength{\itemsep}{-2pt}
\item accept input with equations in abstract index notation;
\item generate customized finite differencing operators;
\item generate codes compatible with advanced Cactus features such as
  adaptive mesh refinement or multi-block systems;
\item check the consistency with non-Kranc generated parts of the
  user's simulation;
\item apply coordinate transformations, in particular of derivative
  operators, suitable for multi-block systems
  (e.g.~\cite{Pollney:2009yz});
\item use symbolic algebra based on the high-level description of the
  physics system to perform optimizations that are inaccessible to the
  compiler of a low-level language;
\item implement transparent OpenMP parallelization;
\item explicitly vectorize loops for SIMD architectures (using
  compiler-specific syntaxes);
\item generate OpenCL code (even independent of the CaKernel framework
  described below);
\item apply various transformations and optimizations (e.g.\ loop
  blocking, loop fission, multi-threading, loop unrolling) as
  necessary for the target architecture.
\end{itemize}

\subsubsection{Optimization}

It is important to note that Kranc does not simply generate the source code for
a specific architecture that
corresponds $1:1$ to its input. Kranc has many of the features of a traditional compiler, including
a front-end, optimizer, and code generator, but the code generated is C++/CaKernel/CUDA rather than
machine code.

The high-level optimizations currently implemented act on discretized
systems of equations, and include the following:
\begin{itemize}
\setlength{\itemsep}{-2pt}
\item Removing unused variables and expressions;
\item Transforming expressions to a normal form according to
  mathematical equivalences and performing \emph{constant folding};
\item
    Introducing temporaries to perform \emph{common subexpression elimination};
\item Splitting calculations into several independent calculations
  to reduce the instruction cache footprint and data cache pressure
  \emph{(loop fission)};
\item Splitting calculations into two, the first evaluating all
  derivative operators (using stencils) storing the result into
  arrays, the second evaluating the actual RHS terms but not using any
  stencils. This allows different loop optimizations to be applied to
  each calculation, but requires more memory bandwidth \emph{(loop
    fission)}.
\end{itemize}

Note in the above that a
\emph{calculation} is applied to all grid points, and thus either
loops over or uses multiple threads to traverse all grid points.
Also note that both the high-level and the low-level optimizations could in principle
also be performed by an optimizing compiler. However, none of the
currently available compilers for HPC systems are able to do so,
except for very simple kernels. We surmise that the reason for this is
that it is very difficult for a compiler to abstract out sufficient
high-level information from code written in low-level languages
to prove that these
transformations are allowed by the language standard. A programmer is
forced to make many (ad-hoc) decisions when implementing a system of
equations in a low-level language such as C or C++, and the compiler
is then unable to revert these decisions and fails to optimize the
code.

It is surprising to see that these optimizations -- which are in
principle standard transformations among compiler builders -- are (1)
able to significantly improve performance, are (2) nevertheless not
applied by current optimizing compilers, and are yet (3) so easily
implemented in Mathematica's language, often requiring less than a
hundred lines of code.

Kranc is a developed and mature package. Since its conception in 2002,
it has been continually developed to adapt to changing computational
paradigms.
Kranc is not just a theoretical tool. In the Einstein
Toolkit~\cite{Loffler:2011ay},
Kranc is used to generate a highly efficient
open-source implementation of the Einstein equations as well as
several analysis modules.
All of the above features are used heavily by users of the Toolkit,
and hence have been well-tested on many production architectures,
including most systems at NERSC or in XSEDE\@.

\subsubsection{Debugging the Numerical Code}
It is also important to note that Chemora significantly reduces the time
required to debug the application.  The recommended approach for development
using Chemora is that the user's Kranc script is considered the canonical
source, and only this should be modified during development. The generated code
should not be modified, as it will be completely regenerated each time Kranc is
run, so any hand-modifications of the generated code will be lost. Unlike when
writing a C++ program, every successfully-compiled Kranc script should lead to
correct computational (though not necessarily physical) code. Hence the errors
are limited to the application domain, for example an incorrect equation is solved.
Similarly, use of a source-code level debugger is
not typical when working with Kranc, as the ``debugging'' happens at the level of
the scientific results (e.g. convergence tests and visualisation) rather than
at the level of programmatic bugs in the generated code. 
As such, Kranc is treated as a black box by the application scientist,
much as a compiler would be.

\subsubsection{Code Generation for CaKernel}

In order to use Kranc as a component of \proj, the code-generation
backend was modified, and
CaKernel (see section \ref{sec:cakernel}) was added as an output target. This
change is essentially invisible to the application developer; there is merely
an additional option to generate CaKernel code rather than C++ or OpenCL code.
Each calculation is then annotated with whether
it runs on the host (CPU) or the device (GPU\@).  Kranc
also creates all metadata required by CaKernel.
Additionally, the new EDL language frontend was added to Kranc.

\subsubsection{Hybrid Codes}

Since GPU accelerators have to be governed by CPU(s), it is natural to attempt
to exploit them by employing \emph{hybrid codes}. In this case, Kranc,
generates both CPU and CaKernel codes from the same script. At
run time, each MPI process checks whether to attach itself to a GPU
and perform its calculations there, or whether to use the CPU for
calculations. 

This mechanism works in principle; however, as the Cactus driver
currently assigns the same amount of work to each
MPI process (uniform load balancing), the large performance disparity between
CPU and GPU has led to only minimal performance gains so far. We expect this issue
to be resolved soon.

\subsection{CaKernel GPU Code Optimization}

\bitbucket{
\proj optimizes at multiple levels of abstraction, starting with the
Kranc scripts, through domain decomposition and data staging,
GPU code generation, down to execution configuration and cache
settings. Optimizations are performed at the appropriate level, but
take advantage of information provided by higher levels.}

\bitbucket{
One goal of \proj is to generate efficient code starting from a
high-level description and without requiring the user to tune for
efficient execution. CaKernel achieves this by using
Kranc-provided and run-time information to set such important
parameters as tile shape and GPU cache settings.}

The CaKernel code generated by Kranc consists of
\term{numerical kernels}, routines that operate on a single grid
point. The CaKernel parts of \proj use Kranc-provided and
run time information to generate efficient GPU executables
from the numerical kernels, without requiring the user to set tuning
parameters. At build time, numerical kernels are wrapped with
\term{kernel frames}, code that implements data staging and iteration,
producing a source code package that is compressed and compiled into
the Cactus executable. At run time, CaKernel makes use of
information about the kernels provided by Kranc as well as user
parameters and information on the problem size to choose tiling, etc.
With this information, the code package is extracted,
edited, compiled, loaded to the GPU, and run. This dynamic process
results in lightweight GPU code that makes efficient use of GPU
resources, including caches.
CaKernel uses several techniques to generate efficient GPU code which
we shall elaborate in the following subsections.

\subsubsection{Stencils and Dynamic Tile Selection}

\bitbucket{
Classic CPU loop tiling involves choosing an iteration strategy to maximize
cache reuse, see for example \cite{rivera00}. Cache reuse is important
for GPU tiling too, however because of latency hiding with
multithreading, a primary goal can be minimization of the total data
request size. Many tiling strategies for stencil computations on GPUs
have been reported~\cite{renganarayana07,datta08,meng09,unat11}, the
common goal being to make best use of the limited amount of high-speed
memory by taking advantage of the repeated access to data elements.}

CPU and GPU tiling has been extensively studied, though often limited
to specific stencils,
\cite{renganarayana07,datta08,meng09,unat11}. The goal for CaKernel
was to develop an automatic tile selection scheme that would work well
not just for a few specific stencils, but any stencil pattern the user
requested. The tile selection is based not just on the stencil shape
but also on the number of grid variables and on the shape of the local
grid. The resulting tile makes best use of the cache and potentially
registers for minimizing data access. The discussion below provides
highlights of the scheme; details will be more fully reported
elsewhere.

The following discussion uses CUDA terminology, see \cite{cuda40,cudatune40}
for background. The term \term{tile} will be used here to mean the portion of
the grid assigned to a CUDA block. In GPUs, higher \term{warp} occupancy means
better latency hiding introduced by common memory access. That can
be achieved with multiple blocks, but to maximize L1 cache reuse
CaKernel will favor a single large block, the maximum block size
determined by a trial compilation of a numerical kernel. Within that
block size limit a set of candidate tile shapes are generated using
simple heuristics, for example, by dividing the $x$ dimension of the
local grid evenly, (by 1, 2, $3,\ldots$) and then for each tile $x$
length find all pairs of $t_y$ and $t_z$ lengths that fit within the
block limit, where $t_x$, $t_y$, and $t_z$ are the tile shape in units
of grid points.

Given a candidate tile shape, the number of cache lines requested
during the execution of the kernel is computed. Such a \term{request
  size} is computed under the \term{ordering assumption} that memory
accesses are grouped by grid function and dimension (for stencil
accesses). As an illustration, if the assumption holds a possible
access pattern for grid functions $g$ and $h$ is $g_{0,1,0}$,
$g_{0,2,0}$, $g_{1,0,0}$, $h_{0,0,0}$, while the pattern $g_{0,1,0}$,
$h_{0,0,0}$, $g_{1,0,0}$, $g_{0,2,0}$ violates the assumption because
$h$ is between $g$'s accesses and for $g$ a dimension-$x$ stencil
access interrupts dimension-$y$ accesses.

Request sizes are computed under different cache line \term{survival
  assumptions}, and the one or two that most closely match the cache
are averaged. One survival assumption is that all lines survive (no
line is evicted) during an iteration in which case the request size is
the number of distinct lines the kernel will touch, after accounting
for many special cases such as alignment. Another survival assumption
is that data accessed using stencils along one dimension (say, $x$)
will not survive until another dimension access (say, $y$)
  (e.g., common lines might be evicted). The particular assumption to
use is based on the size of the tile and cache.

Skipping details, let $r$ denote the overall request size. An
\term{estimated cost} is computed by first normalizing $r$ to the number of
grid points, $r/It_xt_yt_z$, where $I$ is the number of iterations
performed by threads in the tile. To account for the lower execution
efficiency with smaller tiles, a factor determined empirically as
$1/(1+256/t_xt_yt_z)$ is used. The complete expression for the
estimated cost
is $\sigma=(r/It_xt_yt_z)/(1+256/t_xt_yt_z)$.
The tile with the lowest estimated cost is selected.

Tiles chosen using this method are often much longer in the $x$
direction than other dimensions, because the request size includes the
effect of partially used cache lines. If a stencil extends in all three
dimensions and there are many grid functions, the tile chosen will be
``blocky''. If there are fewer grid functions, the tile will be
plate-shaped, since the request size accounts for cache lines that survive
iterations in the axis orthogonal to the plate. The tile optimization
is performed for the tile shape, but not for the number of iterations
which so far is chosen empirically.

\subsubsection{Lightweight Kernel Generation}

A number of techniques are employed to minimize the size of the GPU
kernels. Dynamic compilation using program parameters and tile shape,
seen by the compiler as constants, was very effective. Another
particularly useful optimization given the large size of the numerical
kernels is \term{fixed-offset loads}, in which a single base address
is used for all grid functions. Normally, the compiler reserves two
32-bit registers for the base address of each grid function, and uses
two additional registers when performing index arithmetic since the
overhead for indexing is significant. Fortunately, the Fermi memory
instructions have a particularly large offset, at least 26 bits based
on an inspection of Fermi machine code (which is still not well
documented). (An offset is a constant stored in a memory instruction,
it is added to a base address to compute the memory access address.)
With such generous offsets, it is possible to treat all grid
functions (of the same data type) as belonging to one large array.

\subsubsection{Fat Kernel Detection}

Some numerical kernels are extremely large, and perform very poorly
using standard techniques, primarily due to very frequent register
spill/reload accesses. 
CaKernel identifies and provides
special treatment for such kernels. The kernels can be automatically
identified using CaKernel's integrated performance monitoring code by
examining the number of local cache misses. (Currently, they are
automatically identified by examining factors such as the number
of grid functions.)
Such fat kernels are handled using two techniques: they are launched
in small blocks of 128 threads, and source-level code restructuring
techniques are applied. Launching in small blocks relieves some
pressure on the L1 cache. (A dummy shared memory request prevents
other blocks from sharing the multiprocessor.) The source code
restructuring rearranges source lines to minimize the number of live
variables; it also assigns certain variables to shared memory.

\subsubsection{Integrated Performance Monitoring}

CaKernel provides performance monitoring using GPU event counters,
read using the NVIDIA Cupti API\@. If this option is selected, a
report on each kernel is printed at the end of the run. The report
shows the standard tuning information, such as warp occupancy and
execution time, and also cache performance data. To provide some insight
for how well the code is performing, the percentage of potential
instruction execution and memory bandwidth used by the kernel is
output. For example, a 90\% instruction execution potential would
indicate that the kernel is close to being instruction bound. We plan
to use these data for automatic tuning, e.g.\ to
better identify fat kernels.

\subsubsection{Effectiveness of Low-Level Optimizations}

Most of the optimizations are highly effective, including dynamic
compilation and fixed-offset loads.
There are two areas where some
potential has been left unexploited: tile shape, and the handling of
fat kernels. 

Automatic tile size selection greatly improves performance over
manually chosen tile sizes, however kernels are still running at just
20\% of execution utilization while exceeding 50\% of available memory
bandwidth, suffering L1 cache miss ratios well above what was
expected. The primary weakness in tile selection is assuming an
ordering of memory accesses that does not match what the compiler
actually generates. (The compiler used was NVIDIA \code{ptxas} release
4.1 V0.2.1221.) For example, for a kernel with a $5\times5\times5$
stencil and a $102\times3\times3$ tile, the compiler interleaves $n$
accesses along the $y$ and $z$ axes. The cache can hold all grid
points along one axis (273 cache lines would be needed in this
example) but not along two (483 cache lines). Several solutions have
been identified, including modifying the model to match compiler
behavior, waiting for a better compiler, restructuring the code to
obtain a
better layout, or rescheduling the loads at the object-file level.

One of the kernels performing the last step in the time evolution
has over 800 floating point
instructions in straight-line code. This executes at only 14\%
instruction utilization, suffering primarily from L1 cache misses on
register spill/reload accesses. We address this via fixed offsets and
other dynamic compilation techniques that reduce register pressure. A
combination of source-level scheduling and shared memory use yielded
from 5\% to 10\% better performance, and there seems to be a large
potential for further improvement.

\subsection{Accelerator Framework}
\label{sec:accelerator}

In large, complex applications based on component frameworks such as
Cactus, GPUs and other accelerators are only useful to those
components which perform highly parallel arithmetic computations. As
such, it is neither necessary nor useful to port the entire framework
to run on GPUs -- in fact, much of the code in Cactus-based
applications is not numerical, but provides support in the form of
organizing the numerical data.

One approach to porting a component to run on a GPU is to identify the
entry and exit points of that component, copy all required data to the
GPU beforehand, and copy it back after the GPU computation.
Unfortunately, such data transfer is prohibitively slow, and 
the performance of this approach is not acceptable.

Instead, we track which data (and which parts of the data) is read and
written by a particular routine, and where this routine executes (host
or GPU). Data is copied only when necessary, and then only those
portions that are needed. Note that data is not only accessed for computations,
but also by inter-process synchronization and I/O.

The metadata available for each Cactus component (or thorn) already contains
sufficient information in its schedule description for such tracking,
and during \proj we refined the respective declarations
to further increase performance. This metadata needs to be provided
manually for hand-written thorns, but can be deduced automatically
e.g.\ by Kranc in auto-generated thorns.

In keeping with the Cactus spirit, it is a Cactus component (thorn
\term{Accelerator}) that tracks which parts of what grid functions are
valid where, and which triggers the necessary host--device copy
operations that are provided by other, architecture-specific thorns.

\section{Case studies}

\subsection{Computing Resources}
We tested our framework on different computational systems.
Unfortunately, clusters available to us at the time this paper was written were
insufficient for the purpose of challenging scaling tests.
\todo{Steve: Maybe we can we add Tienhe-1 and/or Super Mike?}

\subsubsection{Cane}
\label{sec:canedesc}

\emph{Cane} is a heterogeneous cluster located at the Pozna\'{n} Supercomputing
and Networking Center.  Although it consists of 334 nodes, at the time we
performed the tests only 40 of them were available as
the cluster was still being set up.
Each node is equipped with two AMD Opteron™ 6234 2.7GHz processors (with two
NUMA nodes each; 12 cores per CPU), 64GB of main memory,
and one NVIDIA M2050 GPU with 3GB of RAM\@. The computational nodes
are interconnected by InfiniBand QDR network with the fat-tree topology
(32Gbit/s bandwidth). CUDA 4.1 and gcc 4.4.5 were used for GPU and CPU code
compilation, respectively.

\subsubsection{Datura}
\label{sec:daturadesc}

\emph{Datura} is an CPU-only cluster at the Albert-Einstein-Institute in
Potsdam, Germany. Datura has 200 nodes, each consisting of two Intel
Westmere 2.666GHz processors with 6 cores and 24GB of memory.
The nodes are connected via QDR InfiniBand (40Gbit/s bandwidth).
We used the Intel compilers version 11.1.0.72.

\subsection{CFD with \proj and Physis}

We employed a simple CFD (Computational Fluid Dynamics) benchmark
application to compare the performance of \proj and Physis.
This code solves the Navier-Stokes equations;
for details about the problem and
its discretization see~\cite{VOF_Hirt79,NASA_VOF2D_Torrey}, and for its
implementation in Cactus and CaKernel see~\cite{parco11,sciprog11,ppopp11}.
The setup consists of three stencil kernels:
one that explicitly updates velocity values, one that iteratively solves 
the conservation of mass (updating velocity and pressure), and one that
updates the boundary conditions.
For simplicity, we ran 4 iterations of the mass conservation kernel,
and applied the boundary condition after each iteration. Although the CFD code
was written directly in CaKernel native language and its performance was
already reported along with our previous
work~\cite{parco11,sciprog11,ppopp11}, we used
CaKernel's new optimization facilities in this work.
These allowed us to obtain improved
performance compared to our previous results as well as compared to
similar, publicly
available frameworks (e.g.\ Physis).

To obtain statistically stable performance results, as many as 1000 iterations
were executed in each run. The CFD benchmark uses single-precision
floating-point data, which provides sufficient accuracy for this test case.
Both frameworks use the GPUs only for computation, and use CPUs only for
data transfer and management.

\begin{figure*}
\centering
\includegraphics[width=0.7\textwidth]{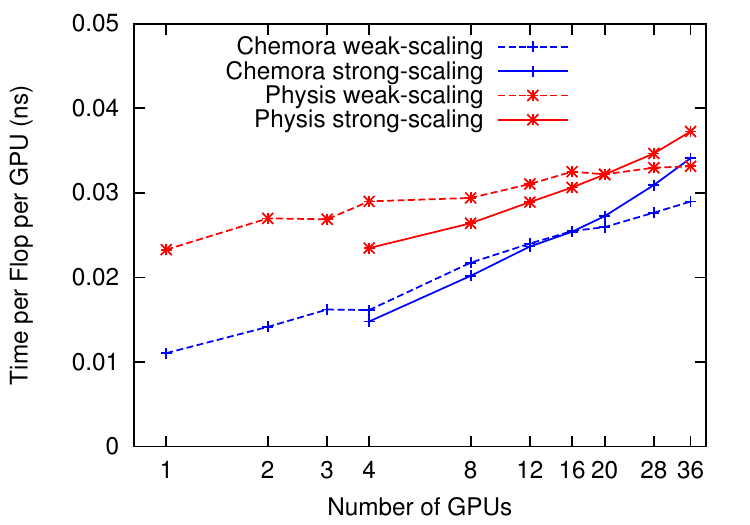}
\caption{Weak and strong scaling test comparing \proj and Physis
  running on multiple nodes for the same CFD application.
  Smaller numbers are better, and ideal scaling correspoonds to a
  horizontal line.
  \proj
  achieves a higher per-GPU performance, whereas Physis shows a higher
  strong scalability. Details in the main text.}
\label{fig:sc12bench_cfd}
\end{figure*}

Figure~\ref{fig:sc12bench_cfd} compares the scalability of the 
frameworks in this CFD benchmark. The problem size of the weak scaling test
for each GPU was fixed at $256^3$, and 
the performance was evaluated using 1 to 36 GPUs with two-dimensional
domain decompositions
along the $y$ and $z$ directions. We present results for
the best domain decompositions for each framework. The performance of both implementations increases
significantly with increasing number of the GPU nodes. Numerous optimizations in \proj such as dynamic 
compilation and auto-tuning allowed us to find the best GPU block size for
the domain size, and execute on the correct number of warps to limit the number of L1 cache 
misses. As a result, for a single GPU, \proj obtained 90.5
GFlop/s, whereas Physis only obtained 43 GFlop/s.
This gap may be also due to the fact that Physis does not make any use of
shared memory on the GPUs.

Figure~\ref{fig:sc12bench_cfd} also compares
the performance of the two frameworks in a strong scaling test.
The problem size for this test was fixed at $656^3$. Both implementations
scale up very well;
\proj achieved 270 GFlop/s and 1055 GFlop/s for 4 and 36 GPUs, respectively, 
whereas Physis achieved 170 GFlop/s and 965 GFlop/s in the same configurations.
The parallel efficiency (when increasing the number of GPUs from 
4 to 36) is 43\% and 63\% for \proj and Physis, respectively.

\subsection{Binary Black Hole Simulations with \proj}

We demonstrate the integration of \proj technologies into our
production-level codes by performing a  Numerical Relativity (NR)
simulation. This simulation of a binary black
hole (BBH) merger event shows that our GPU-accelerated main evolution
code can be seamlessly integrated into the pre-existing CPU framework, and that
it is not necessary to port the entire framework to the GPU\@.  It also
demonstrates the use of the data management aspect of \proj, showing
how data is copied between the host and the device on demand.
Analysis modules
running on the CPU can make use of data generated on the GPU
without significant modification.

Our production simulations differ from this demonstration only in
their use of adaptive mesh refinement (AMR), which allows a much larger
computational domain for a given computational cost.  This allows the
simulation of black hole binaries with larger separations, many more
orbits before merger, and hence longer waveforms when AMR is used.

The initial condition consists of two black holes on a quasi-circular
orbit about their common center of mass (``QC-0'' configuration). This
is a benchmark configuration; in a production simulation, the black
holes would have a much larger separation.
This configuration performs approximately one orbit before the energy
loss due to gravitational wave emission cause the black holes to
plunge together and form a single, highly-spinning black hole.

Gravitational waves are emitted from the orbiting and merging system.
These are evaluated on a sphere and decomposed into spherical
harmonics.
It is this waveform which is used in gravitational wave detection.

We use a 3D Cartesian numerical grid $x^i \in [-6.75, 6.75]^3$ with
$270^3$ evolved grid points. To ensure a balanced domain decomposition
we run on 27 processes, corresponding to $90^3$ evolved points per
process. This is the largest grid that fits in the 3 GB of GPU memory
on Cane, given the large number of grid variables required.
All calculations are performed in double precision.
We evolve the system using the \code{McLachlan} code (see section
\ref{sec:science} above), using 8th order finite differencing and a
3rd order Runge-Kutta time integrator.

Any production Cactus simulation makes use of a large number of
coupled thorns; e.g.\ this simulation contains 42 thorns. Most of
these do not need to be aware of the GPU, CaKernel, or the Accelerator
infrastructure. In our case, only \code{McLachlan} and the
\code{WeylScal4} gravitational wave extraction thorns were running on a
GPU\@. Additional thorns, e.g.\ tracking the location or shape of the
black holes, were run on the CPU\@.

We use 27 nodes of the Cane cluster (see section~\ref{sec:canedesc})
with one GPU per node. We do not run any CPU-only processes.

\begin{figure}
\centering
\includegraphics[width=0.50\textwidth]{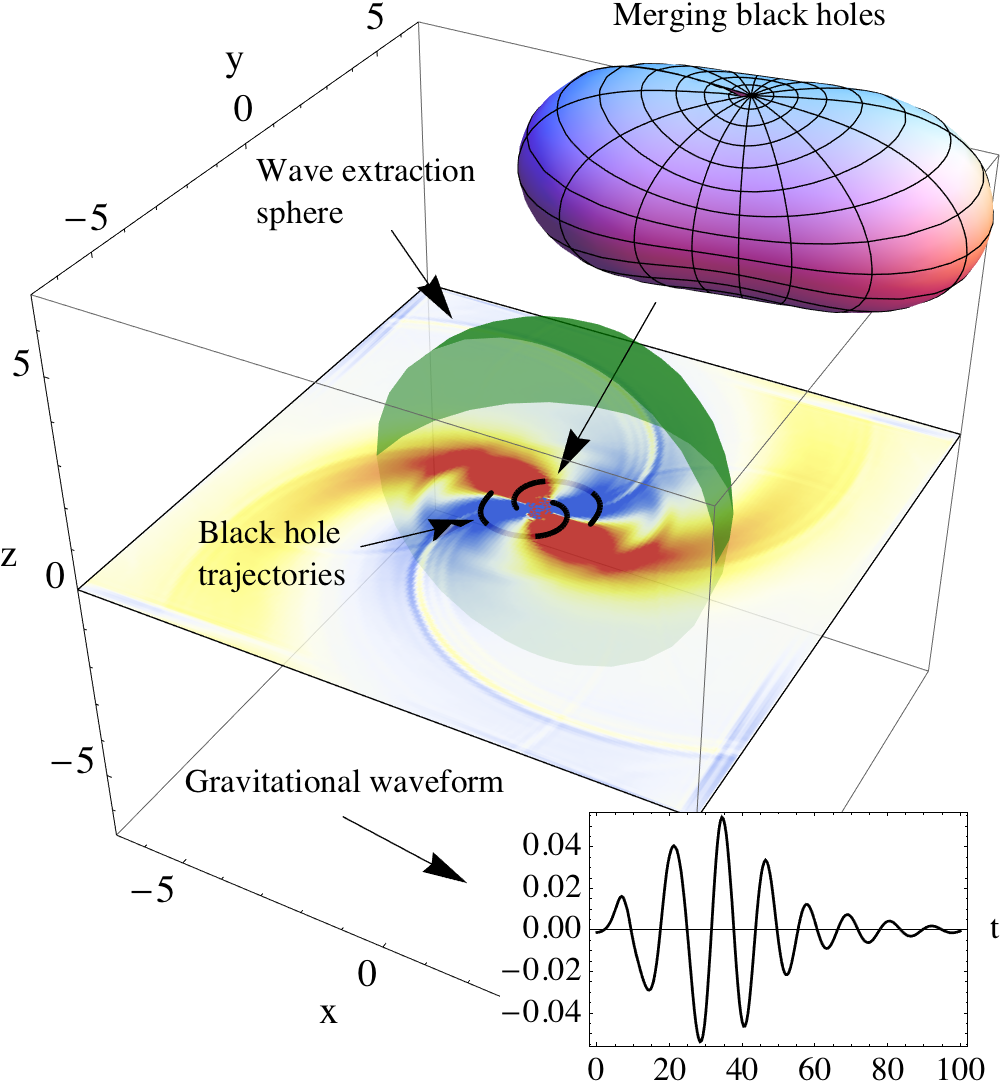}
\caption{Visualization of a binary black hole system}
\label{fig:bbh}
\end{figure}

Fig.~\ref{fig:bbh} shows the numerical simulation domain.  On the
$x-y$ plane we project the $\Psi_4$ variable which represents
gravitational waves.  The black hole trajectories are shown as black
curves near the center of the grid; they end when the black holes
merge into a single black hole located at the center.
The sphere on which
the multipolar decomposition of the gravitational waves is performed
is also shown.  In the insets, we show (a) the time evolution of the
(dominant) $\ell = 2, m = 2$ mode
of the gravitational radiation computed on the sphere at $r = 4 M$,
and (b) the (highly distorted) shape of the common apparent horizon
formed when
the two individual black holes merge.

Table~\ref{tbl:bbhtimers} shows a break-down of the total run time of
the BBH simulation. The routines labeled in bold face run on the
GPU\@. The times measured are averaged across all processes. The
\emph{Wait} timer measures the time processes wait on each other
before an interprocessor synchronization. This encapsulates the
variance across processes for the non-communicating routines.

\begin{table}
\centering
\begin{tabular}{lr}
        & \hspace{-0.5cm}Percentage of \\
  Timer & \hspace{-0.5cm}total evolution time \\
\hline
 \text{Interprocess synchronization} & 39\%\\
 \textbf{RHS advection} & 13\%\\
 \textbf{RHS evaluations} & 12\%\\
 \text{Wait} & 11\%\\
 \textbf{RHS derivatives} & 6\%\\
 \textbf{Compute Psi4} & 5\%\\
 \text{Multipolar decomposition} & 3\%\\
 \text{File output} & 3\%\\
 \text{BH tracking} & 3\%\\
 \text{Time integrator data copy} & 2\%\\
 \text{Horizon search} & 2\%\\
 \textbf{Boundary condition} & 1\%\\
 \text{BH tracking (data copy)} & 1\%\\
\end{tabular}
\caption{Timer breakdown for the binary black hole
  simulation.  Routines in bold face (48\%) are executed on the GPU\@.}
\label{tbl:bbhtimers}
\end{table}

We see that the interprocess synchronization is a significant portion
(38\%) of the total run time on this cluster. One reason for this is
that the large number of ghost zones (5) needed for partially-upwinded 8th order stencils
require transmitting a large amount of data. This could likely be
improved by using a cluster with more than one GPU or more GPU memory
per node, as this would reduce the relative cost of inter-process
communication relative to computation.

\subsection{McLachlan Benchmark}

\begin{figure*}
\centering
\includegraphics[width=0.7\textwidth]{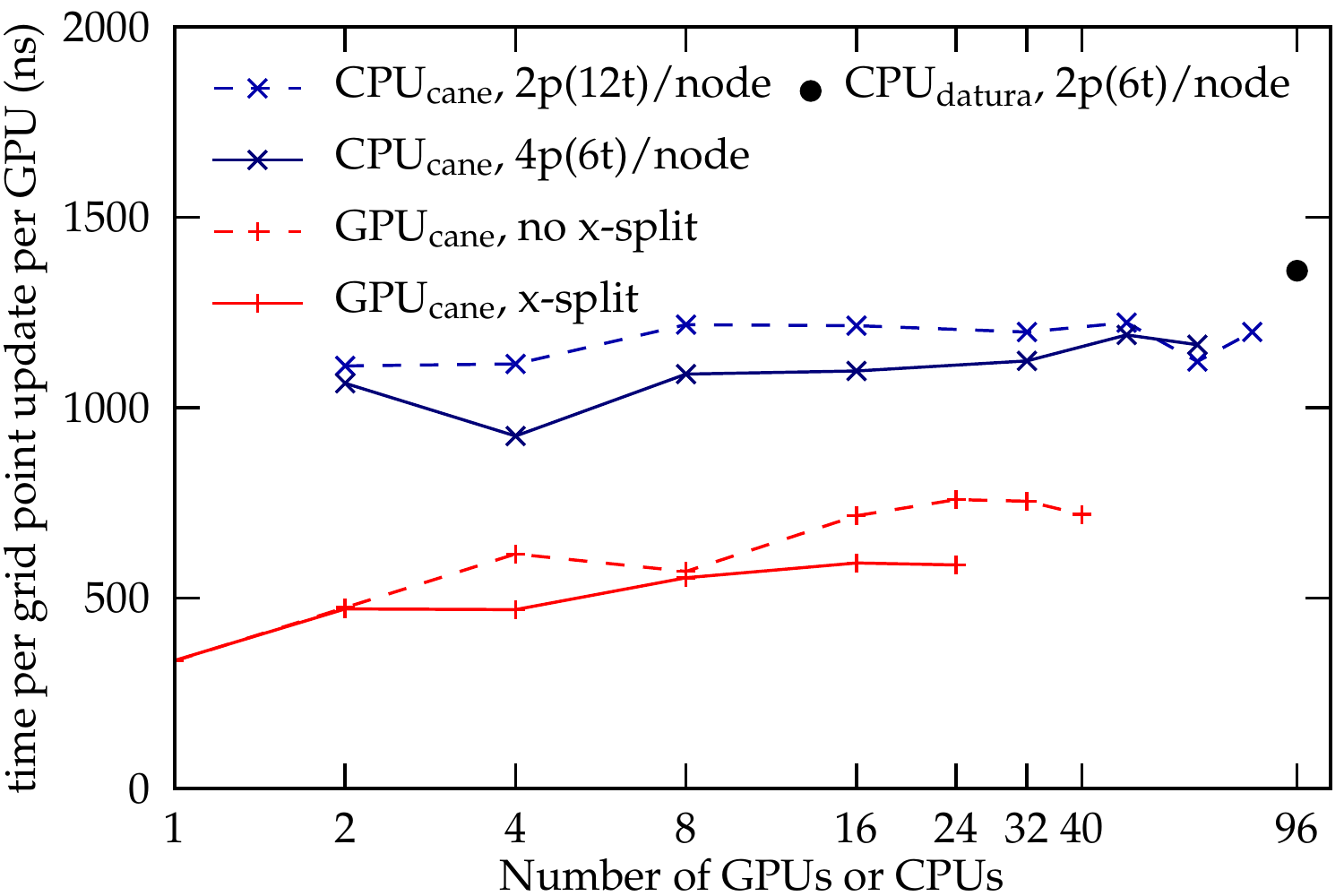}
\caption{Weak-scaling test for McLachlan code performed on the Cane and
  Datura clusters. ($n$)p(($m$)t) stands for $n$ processes per node
  using $m$ threads each. (no) x-split stands for (not) dividing
  domain along the $x$ axis. Smaller numbers are better, and ideal
  weak scaling corresponds to a horizontal line. The benchmark scales
  well on these platforms.}
\label{fig:s12bench_ml}
\end{figure*}

We used part of the binary black hole simulation as a weak-scaling
performance benchmark. We chose a local problem size that fitted into
the GPU memory of Cane (see section~\ref{sec:canedesc}), corresponding
to $100^3$ evolved points plus boundary and ghost zones. We ran
the benchmark on Cane (on GPUs) and Datura (on CPUs; see
Sec.~\ref{sec:daturadesc}), using between 1 and 48 nodes.
Figure~\ref{fig:s12bench_ml} shows results comparing several
configurations, demonstrating good parallel scalability for these core
counts. One of Cane's GPUs achieved about twice the performance
of one of its CPUs, counting each NUMA node as a single
CPU\@.

As a measurement unit we use time per grid point update per GPU (or
CPU\@). The best performance was achieved for a single GPU: 25
GFlop/s, which is 5\% of the M2050 GPU's peak performance of 515
GFlop/s. On 40 nodes, we observed 50\% scaling efficiency due to
synchronization overhead, and achieved a total performance of 500
GFlop/s.

CPU performance tests were performed on both Cane and Datura. The
total performance of the parallel OpenMP code, properly vectorized, was
similar to the performance of a single GPU, with similar scaling factor.

We note that our floating point operation counts consider only those
operations strictly needed in a sequential physics code, and e.g.~do
not include index calculations or redundant computations introduced by
our parallelization. \todo{Steve: I'm not quite sure what this
paragraph is trying to say, it sounds like we're saying our code
has inefficiencies and we artificially removed their effects in
our results.}

\section{Conclusion}

We have presented the \proj project, a component-based approach to 
making efficient use of current and future accelerator
architectures for high-performance scientific codes.  
Although the examples we present run on the GPU and use CUDA, 
our work is general and will be applied e.g.\ to OpenCL and other approaches in
future work. Using \proj, a scientist can 
describe a problem in terms of a system of PDEs in our Equation
Description Language.
A module for the Cactus framework is then
generated automatically by Kranc for one or more
target architectures.
Kranc applies many optimizations 
at code-generation time, making use of symbolic algebra,
and the resulting source code can then be compiled on a
diverse range of machines (taking advantage of the established
portability of Cactus and the availability of CUDA as a uniform GPU
programming environment).  At run-time, the CUDA code is recompiled
dynamically to enable a range of runtime optimizations.

We have presented two case studies. The first is a 
Computational Fluid Dynamics (CFD) code, and we demonstrated weak scaling
using our infrastructure running on GPUs.  We also used
the Physis framework for this same problem and compared the scaling.
\proj has comparable
or higher performance, a result we attribute to the dynamic
optimizations that we employ.  The second case study is a Numerical
Relativity simulation based on the
McLachlan code, a part of the freely available open-source (GPL)
Einstein Toolkit (ET\@).  McLachlan solves a
significantly more complex set of equations, and integrates with many
other components of the ET\@.  We performed a simulation of a binary
black hole coalescence using the same codes and techniques as we would
currently use in production CPU simulations, with the omission of
Adaptive Mesh Refinement (AMR), which is not yet adapted to \proj.

We plan to implement AMR and multi-block methods next.
AMR and multi-block are implemented in Cactus in a
way which is transparent to the application programmer, hence we
expect that including AMR in \proj will be straightforward using
the Accelerator architecture developed in this work
(which maintains knowledge of which variables are valid on the host
(CPU) and which on the device (GPU)).  As with the time integration, we
will implement only the basic low-level interpolation operators
required for mesh refinement on the GPU, and the existing AMR code
Carpet will marshal the required operations to the device.

With AMR and/or multi-block methods, \proj will be an even more
compelling option for implementing scientific codes, and fields of
science (such as Numerical Relativity) requiring the solution of
complex systems of PDEs will be able to reach a new level of
performance.
Should the specifics of accelerator devices change in the future, the
\proj framework, much of which is general, should be easily adaptable to
the new technology, and codes built with \proj will have a head start in
advancing computational science on the new platform.

\section*{Acknowledgments}

The authors would like to thank Gabrielle Allen and Joel E. Tohline at
the CCT
and Krzysztof Kurowski at PSNC for their vision, encouragement, and 
continuous support to this project.

This work was supported by the UCoMS project under award number MNiSW
(Polish Ministry of Science and Higher Education) Nr 469~1~N~-
USA/2009 in close collaboration with U.S. research institutions
involved in the U.S. Department of Energy (DOE) funded grant under
award number DE-FG02-04ER46136 and the Board of Regents, State of
Louisiana, under contract no.\ DOE/LEQSF(2004-07)
and LEQSF(2009-10)-ENH-TR-14.
This work was also supported by NSF award 0725070 \emph{Blue Waters},
NFS awards 0905046 and 0941653
\emph{PetaCactus}, NSF award 0904015 \emph{CIGR}, 
and NSF award 1010640 \emph{NG-CHC} to Louisiana State University, and
by the DFG grant SFB/Transregio~7 ``Gravitational-Wave Astronomy''.

This work was performed using computational resources of XSEDE
(TG-CCR110029, TG-ASC120003), LONI (loni\_cactus), LSU, and PSNC,
and on the Datura cluster at the AEI\@.

\bibliographystyle{unsrt}
{\footnotesize\bibliography{CCT_CS,marqs,GPU,dmk,references,Cactus}}

\end{document}